\documentclass[aps,prd,12pt,preprint,tightenlines,unsortedaddress,
amsfonts,amssymb,amsmath,byrevtex,showpacs,nofootinbib]{revtex4-1}

\usepackage{amssymb,amsmath}
\usepackage{latexsym}
\usepackage{graphicx}
\usepackage{epstopdf}
\usepackage{color}
\usepackage{xcolor}
\usepackage{graphics}
\usepackage{epsfig,subfigure}
\usepackage{bm}        % for math

\newcounter{mnotecount}[section]
\renewcommand{\themnotecount}{\thesection.\arabic{mnotecount}}
\newcommand{\mnote}[1]%{}
{\protect{\stepcounter{mnotecount}}$^{\mbox{\footnotesize  $%\!\!\!\!\!\!\,
      \bullet$\themnotecount}}$ \marginpar{\raggedright\tiny%\em
    $\!\!\!\!\!\!\,\bullet$\themnotecount: #1} }

\def\ud{\mathrm{d}}

\def\R{\mathbb{R}}

\def\vap{\varpi}

\def\ii{\mathrm{i}}

\def\sfM{\mathsf{M}}
\def\sfMps_{\sfM_{\sigma}^{\vap}}

\def\ud{\mathrm{d}}

\def\bsb{\boldsymbol{\beta}}

\def\hq{\hat q}
\def\hp{\hat p}

\begin{document}

\author{Herv\'{e} Bergeron}
\affiliation{ISMO, UMR 8214 CNRS, Univ Paris-Sud, France}

\author{Ewa Czuchry}
\affiliation{National Centre for Nuclear Research, 00-681
Warszawa, Poland}

\author{Jean-Pierre Gazeau}
\affiliation{APC, Universit\'e Paris Diderot, Sorbonne Paris Cit\'e, 75205 Paris
Cedex 13, France}
\affiliation{Centro Brasileiro de Pesquisas Fisicas
22290-180 - Rio de Janeiro, RJ, Brazil }

\author{Przemys{\l}aw Ma{\l}kiewicz}
\affiliation{National Centre for Nuclear Research,  00-681
Warszawa, Poland}

\date{\today}

\title{Integrable Toda system as a quantum approximation to the anisotropy of Mixmaster}

\begin{abstract}
{We present a regularisation approach to the study of the quantum dynamics of the Mixmaster universe which allows to approximate the anisotropy potential with the explicitly integrable periodic 3-particle Toda system. This approach is based on a covariant Weyl-Heisenberg integral quantization. Such a procedure  naturally amplifies the dynamical role of the underlying Toda system  by smoothing out the three canyons of the anisotropy potential.  Since the respective eigenfunctions can be explicitly constructed, our finding paves the way to a novel perturbative approach to the quantum Mixmaster dynamics.}

\end{abstract}
 \maketitle
%%%98.80.Qc,04.30.Db,04.20.Dw,04.60.Kz

\tableofcontents

\section{Introduction}
 { A widely acknowledged description of the universe on the approach to the cosmological singularity is given by the Belinskii-Khalatnikov-Lifshitz (BKL) scenario \cite{BKL,cwm}, generalised and improved in \cite{Uggla1,Uggla2}. It asserts that on the approach to the singularity the isotropy and homogeneity of space-time break down \cite{Lifshitz}. However, it also shows at each spatial point asymptotically towards the singularity the dynamics would approximate to a spatially  anisotropic but homogenous model.  in particular the Bianchi type VIII or IX \cite{Uggla1,Uggla2}. The spatially homogeneous models of general relativity are classified into the so-called Bianchi types (see e.g. \cite{cosm}) according to the algebra of their three spatial Killing vectors.  The most generic ones are  the Bianchi type VIII or IX. The BKL scenario agrees with numerical studies  which assumes  no space-time   symmetries  and uses convenient scale-invariant Hubble normalized variables of \cite{Uggla1} that reflect the general scale invariance of Einstein equations \cite{DG}. }

{The classical dynamics of the Bianchi IX model,  or mixmaster universe, was  studied by Misner in the canonical formalism\cite{cwm}. It  involves a Hamiltonian that is formally identical to that of a particle moving in the 3D Minkowski spacetime in a time-dependent, exponentially steep and triangle-like potential. The asymptotic classical dynamics of the model is usually expressed by means of the so called Kasner map that consists of an infinite sequence of Kasner universes \cite{BKL,cwm, Uggla1,Uggla2}. At each transition between two consecutive Kasner epoch, the Misner particle bounces off the steep potential. Many mathematical studies have been devoted to the classical evolution of the mixmaster universe and have led to some important results on the asymptotic behaviour, the non-integrability or the chaotic behaviour (see e.g. \cite{Uggla1,Uggla2}, \cite{Corn}). } 

{On the other hand, the quantum dynamics of mixmaster remains poorly understood despite many interesting studies (see e.g. \cite{cwm, marolf, moncrief, damour} or a review \cite{Montani}). A recent result \cite{vibronic} shows that the problem can be considerably simplified if one uses the so called vibronic approach to the coupling between spatial (i.e. anisotropic) and  temporal (i.e. isotropic) variables. The difficulty now lies in the elaborate spatial dependence of the anisotropy potential and the fact that solving the quantum dynamics requires the knowledge of the eigenstates and the eigenvalues for the anisotropic motion. The solution to the eigenvalue problem seems to  have been so far approximated by means of the harmonic and the steep-wall approximation to the anisotropy potential (see e.g. \cite{qb9f} for their proper derivation).  They correspond to the large volume and low anisotropy excitation level or  the small volume and high anisotropy excitation level, respectively. However, it was shown \cite{vibronic,inflation} that the existence of a quantum bounce, a generic prediction of successful quantum models, induces a nonadiabatic evolution to which neither of these approximations is applicable. It seems therefore crucial to find a more suitable approximation. }

{In this paper we present a new approach to the quantum mixmaster dynamics within which we derive and justify an appealing quantum approximation to the anisotropy potential.  We employ the so-called Weyl-Heisenberg (W-H) covariant integral quantization \cite{hbjpg2014,becugaro17_1,gazeau_intsympro18}. It is a general procedure that in some circumstances provides conceptually better justified quantization prescriptions than the familiar `canonical prescription', although it includes the latter as a particular case. While respecting  the Weyl-Heisenberg symmetry, our procedure deforms the familiar canonical  quantization by introducing  a  weight function on the classical phase space.   In the present case, the weight is a product of Gaussians whose widths label a family of quantum models. The proposed quantization naturally introduces a `small' perturbation parameter into the system, which emphasizes the role of the Toda potential as the anisotropy potential in the quantum dynamics. In fact, this procedure points to the Toda potential as a manageable approximation to the anisotropy potential.}

{In the simplest case Toda systems} consist of a finite chain of $N$ equal-mass particles with interaction varying exponentially with their separation. They were originally introduced to study analytical solutions of the motion of nonlinear integrable chains \cite{Toda}. The three-particle Toda lattice is the simplest non-trivial integrable nonlinear chain whose integrability was proved a long time ago \cite{Henon,Flaschka,Ford}. {At the quantum level the}  wavefunctions of the corresponding Schr\"odinger equations were rigorously studied \cite{Gutzwiller}  through the symmetries of the system.  {In this approach}, this system {describes} slightly perturbed harmonic chains in the small energy limit, {whereas} in the high energy limit it {becomes} equilateral triangular billiard \cite{Isola}.  Those low and high energy limits {closely resemble the analogous limits of the anisotropic part of the Mixmaster Hamiltonian, recently discussed in \cite{qb9f}.}

{At the classical level, the similarity between the anisotropy potential and the Toda one was used to introduce the so called disturbed Toda lattices \cite{Bogo} that include the anisotropy potential and the Toda one as special cases. Nevertheless, due to the integrability of the Toda latices this similarity seems too far to be useful in the context of the classical dynamics (though see \cite{BSz}). On the other hand, it has never been shown justified and exploited at the quantum level. In the present paper we provide a completely new link between  the anisotropy and the exact Toda potential, which is valid for quantum models. Since  the eigenvalue problem for the Toda lattice can be solved, the proposed approximation should be very useful in future investigations of the quantum dynamics, in particular if combined with the mentioned vibronic approach.}

{The Mixmaster anisotropy potential can be viewed as a difference between  two 3-particle Toda potentials. Our quantization procedure  amplifies the relative contribution of the positive Toda potential  that is shown dominate over the negative one. Thus, the latter can be neglected in the first approximation and} the approximate evolution of the anisotropic variables for a fixed  volume of the universe becomes explicitly integrable. The negative Toda potential can be treated as a perturbation to the integrable dynamics.
 This split opens the door to better understanding of the quantum dynamics,  {essentially} through {perturbative} methods.

The outline of the paper is as follows. In Sec. II we consider the Bianchi IX Hamiltonian constraint, the anisotropy potential, its previously known approximations and the new approximation based on the Toda potential. In Sec. III we explain the basics of the Weyl-Heisenberg {covariant} integral quantization. In Sec. IV we study the  W-H integral {covariant} quantization of the anisotropy potential and show that the proposed approximation is reaffirmed at the quantum level. We conclude in Sec. V.

\section{Decomposition of the {classical} anisotropy potential}\subsection{Definition of the model}
\begin{figure}[t]
\includegraphics[width=0.35\textwidth]{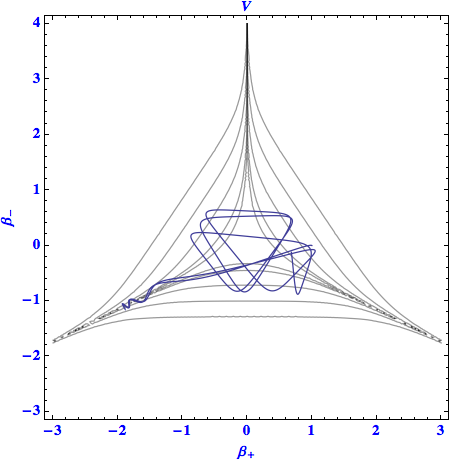}
\caption{{A sample   solution to the motion of a particle in the anisotropy potential}. The contours  are $3, 10^1, 10^2, 10^3, 10^4$ and the initial condition for the plotted solution is $p_+=10$, $p_-=-10$, $\beta_+=0$, $\beta_-=1$.}
\label{figure1}
\end{figure}

Let us first recall the definition and the Hamiltonian formulation of the Bianchi type IX model. We assume the line element:
\begin{equation}
\ud s^2= -{\cal N}^2\ud\tau^2+\sum_ia_i^2(\omega^i)^2\, ,
\end{equation}
where {$\ud \omega_i=\frac{1}{2} \frak{n}\varepsilon_{i}^{\, jk}\omega_j \wedge \omega_k$ and ${\cal N}$, $a_i$ are functions of time. The respective Hamiltonian constraint in the Misner variables reads \cite{cwm}:
\begin{equation}\label{con}
\mathrm{C}=\frac{{\cal N}e^{-3\Omega}}{24}\left(\frac{2\kappa}{\mathcal{V}_0}\right)^2\left(-p_{\Omega}^2+\mathbf{ p}^2+36\left(\frac{\mathcal{V}_0}{2\kappa}\right)^3\frak{n}^2e^{4\Omega}[V(\bsb)-1]\right)\,,~~(\Omega,p_{\Omega},\bsb, \mathbf{p})\in\mathbb{R}^6,
\end{equation}
where $\bsb:=(\beta_{+},\beta_-)$, $\mathbf{p}:= (p_+,p_-)$, $\mathcal{V}_0=\frac{16\pi^2}{\frak{n}^3}$ is the fiducial volume, $\kappa=8\pi G$ is the gravitational constant, ${\cal N}$ is the non-vanishing and otherwise arbitrary lapse function. In what follows we set  $\frak{n}=1$ and $2\kappa=\mathcal{V}_0$. The gravitational Hamiltonian $\mathrm{C}$ resembles the Hamiltonian of a particle in the 3D Minkowski spacetime in a  {certain} potential. The spacetime variables used in Eq. \eqref{con} have the cosmological interpretation:
\begin{equation}\Omega=\frac13\ln a_1a_2a_3,~~\beta_+=\frac16\ln\frac{a_1a_2}{a_3^2},~~\beta_-=\frac{1}{2\sqrt{3}}\ln\frac{a_1}{a_2}~.\end{equation}
Thus, the variable $\Omega$ describes the isotropic geometry, whereas $\beta_{\pm}$ describe distortions to the isotropic geometry and are referred to as the anisotropic variables. The potential that drives the motion of the geometry represents the spatial curvature $^3R$.

The Hamiltonian constraint (\ref{con}) is a sum of the isotropic and anisotropic parts, $\mathrm{C}=-\mathrm{C}_{iso}+\mathrm{C}_{ani}$, where (up to a factor)
\begin{align}\label{condec1}
\mathrm{C}_{iso}&=p_{\Omega}^2+36e^{4\Omega},\\
\label{condec2}\mathrm{C}_{ani}&=\mathbf{p}^2+36e^{4\Omega}V(\bsb)\, ,
\end{align}
and where
\begin{equation}\label{b9pot}
V(\bsb) = \frac{e^{4\beta_+}}{3} \left[\left(2\cosh(2\sqrt{3}\beta_-)-e^{- 6\beta_+}
\right)^2-4\right] +  1 \,.
\end{equation}
The potential $V(\bsb)$ is plotted in Fig. \ref{figure1}. In the present paper we propose a new approach to the anisotropic Hamiltonian (\ref{condec2}) for a fixed value of the isotropic variable $\Omega$  {(whose value  we} set as  $36e^{4\Omega}=1$)}, specifically we study  \begin{align}\label{focus}
\mathrm{C}_{ani}=\mathbf{p}^2+V(\bsb).
\end{align}

\begin{figure}[t]
\begin{tabular}{cc}
\includegraphics[width=0.35\textwidth]{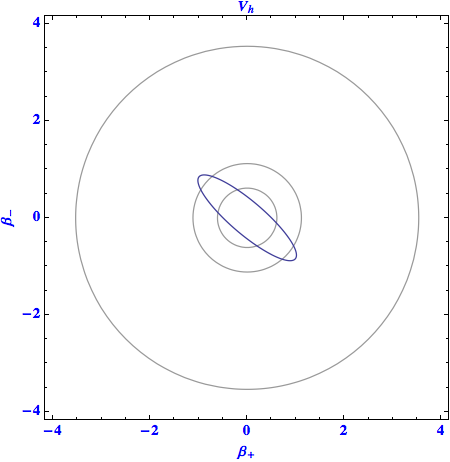}
\includegraphics[width=0.36\textwidth]{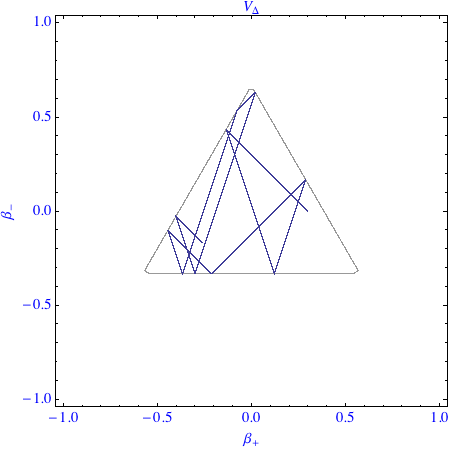}
\end{tabular}
\caption{{Sample solutions of the motion in} the harmonic (on the left) and steep-wall (on the right) approximations to the anisotropy potential. The contours  are $3, 10^1, 10^2$  for the case of the harmonic approximation and the initial condition for the plotted solutions is $p_+=10$, $p_-=-10$, $\beta_+=0$, $\beta_-=0.3$. We notice that the approximate dynamics is very different from the exact one. The harmonic approximation produces a simple ellipsoidal motion, while the steep-wall approximation confines the motion to an arbitrary, fixed triangular potential.}
\label{figure11}
\end{figure}

The issue of {the} coupling between the anisotropic and isotropic variables (at the quantum level) is addressed elsewhere \cite{nonadiabatic}. {The presently known approximations to the anisotropy potential are the harmonic and steep-wall approximation, $V_h(\bsb)$ and $V_{\triangle}(\bsb)$:
\begin{align}
V_h(\bsb)&= 8 (\beta_+^2+\beta_-^2),\\
V_{\triangle}(\bsb)&= \lim_{q\to 0} \frac{1}{3} \, \exp \left[ 4 |\ln q| (\beta_++\sqrt{3} |\beta_-|) \right].
\end{align}}
They are plotted in Fig. \ref{figure11}.

\subsection{Decomposition}

We decompose the anisotropy potential as follows
\begin{align}\nonumber
V(\bsb)&=\frac{1}{3}\left(e^{4\sqrt{3}\beta_-+4\beta_+}+e^{-4\sqrt{3}\beta_-+4\beta_+}+e^{-8\beta_+}\right)\\\label{split}&-\frac{2}{3}\left(e^{-2\sqrt{3}\beta_--2\beta_+}+e^{2\sqrt{3}\beta_--2\beta_+}+e^{4\beta_+}\right)+1\\\nonumber &=V_T+V_{p}+1.
\end{align}
The introduction of new variables $q_1,~q_2,~q_3$ such that $q_1-q_2=4\sqrt{3}\beta_-+4\beta_+$ and $q_2-q_3=-4\sqrt{3}\beta_-+4\beta_+$ leads to
\begin{align}\nonumber
V_T=\frac{1}{3}\left(e^{q_1-q_2}+e^{q_2-q_3}+e^{q_3-q_1}\right),\\
V_p=-\frac{2}{3}\left(e^{-\frac{1}{2}(q_1-q_2)}+e^{-\frac{1}{2}(q_2-q_3)}+e^{-\frac{1}{2}(q_3-q_1)}\right).
\end{align}
This coordinate transformation can be consistently extended by assuming an extra variable, say $\beta_z$, which is absent in the potential,
\begin{align}
\begin{bmatrix}
 \beta_+  \\
 \beta_-  \\
 \beta_z
 \end{bmatrix}
 =
\begin{bmatrix}
 \frac{1}{8} & 0 & - \frac{1}{8} \\
 \frac{1}{8\sqrt{3}} & -\frac{1}{4\sqrt{3}} & \frac{1}{8\sqrt{3}} \\
 a & b & c
 \end{bmatrix}
 \,
 \begin{bmatrix}
 q_1  \\
 q_2  \\
 q_3
 \end{bmatrix},
\end{align}
where $a,~b,~c$ are such that the transformation is invertible. Then we readily obtain the relation between conjugate momenta,
\begin{align}
\begin{bmatrix}
 p_1  \\
 p_2  \\
 p_3
 \end{bmatrix}
 =
\begin{bmatrix}
\frac{1}{8} &  \frac{1}{8\sqrt{3}} & a \\
 0 & -\frac{1}{4\sqrt{3}} & b \\
 - \frac{1}{8} & \frac{1}{8\sqrt{3}} & c
 \end{bmatrix}
 \,
 \begin{bmatrix}
 p_+  \\
 p_-  \\
 p_z
 \end{bmatrix},
\end{align}
and the Hamiltonian
\begin{align}\label{focus2}
\mathrm{H}=32(p_1^2+p_2^2+p_3^2)+V_T+V_p+1,
\end{align}
which yields the anisotropic Hamiltonian (\ref{focus}) for the vanishing total momentum $p_1+p_2+p_3=0$ (or, $p_z=0$). Note that the total momentum is a quantity conserved by (\ref{focus2}).

\begin{figure}[t]
\includegraphics[width=0.30\textwidth]{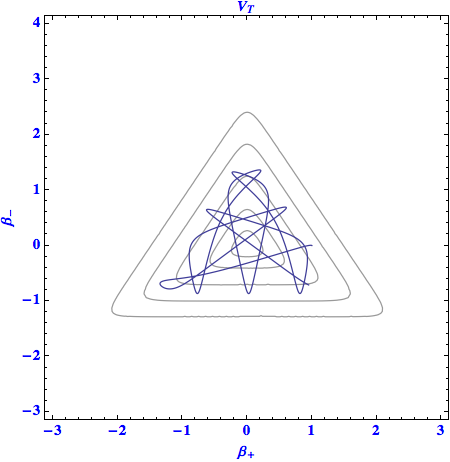}
\caption{Contour plot of the Toda potential and a sample solution. The contours  are $3, 10^1, 10^2, 10^3, 10^4$ and the initial condition for the plotted solution is $p_+=10$, $p_-=-10$, $\beta_+=0$, $\beta_-=0.3$. We notice that the particle's motion resembles the exact motion outside the narrow canyons which are smoothed out in the present approximation.}
\label{figure111}
\end{figure}

\subsection{Perturbation of the Toda system}
Upon rescaling, $q_i\rightarrow \lambda q_i$, $p_i\rightarrow \lambda^{-1} p_i$ and $t\rightarrow 3e^{-\lambda} t$ such that $3e^{-\lambda}=\lambda^{2}$, one brings (\ref{focus2}) to the following form (up to an irrelevant constant):
\begin{align}\label{focus3}
\mathrm{H}=\frac{1}{2}(p_1^2+p_2^2+p_3^2)+e^{q_1-q_2}+e^{q_2-q_3}+e^{q_3-q_1}+3e^{-\frac{1}{2}\lambda}V_p.
\end{align}
The above Hamiltonian describes the periodic 3-particle Toda system \cite{berry76} perturbed by another 3-particle Toda potential $3e^{-\frac{1}{2}\lambda}V_p$.

 {The periodic Toda system is a system of $N$ equal-mass particles interacting via exponential forces, described by the Hamiltonian:
\begin{equation}
H=\frac12\sum_{k=1}^N p_k^2 +\sum_{k=1}^N e^{-(q_k-q_{k+1})}
\end{equation}
with periodicity condition $q_0 \equiv q_N$ and $q_1\equiv q_{N+1}$.}

\begin{figure}[!ht]
\includegraphics[scale=0.15]{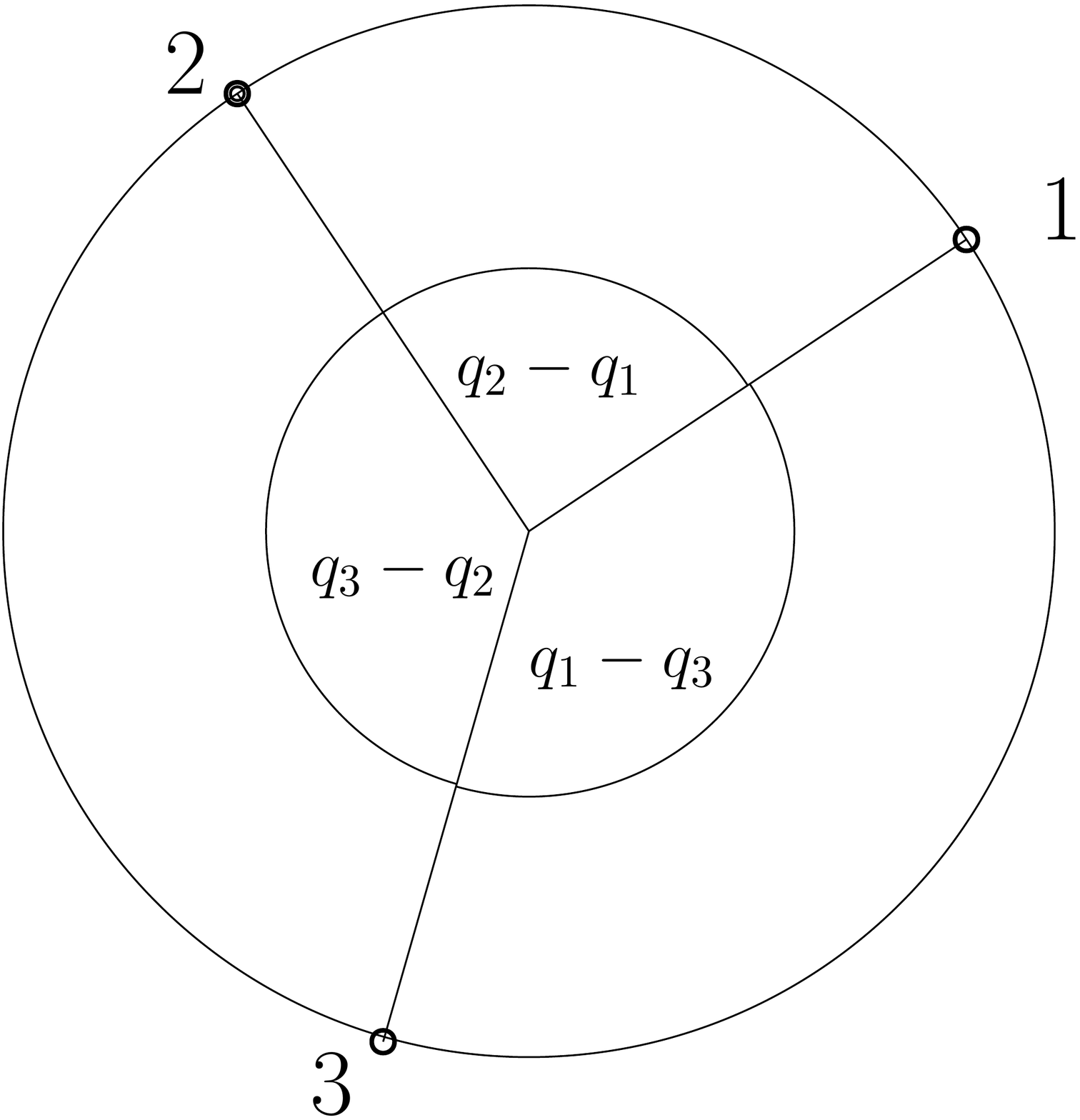}
\caption{Model for the periodic 3-particle Toda lattice. The particles on the lattice  interact with the neighbour ones via exponential potential.}\label{3Toda}
\end{figure}

The periodic 3-particle Toda system is the simplest nontrivial  periodic crystal consisting of three particles, see Fig. \ref{3Toda}. It is known that the Toda systems are integrable {\cite{Henon,Flaschka,Ford}} and solutions can be derived. This system has three independent conserved quantities: the total momentum $P=p_1+p_2+p_3$, the total energy $H$ and an additional third invariant:
\begin{equation}
 K=-p_1 p_2 p_3+ p_1e^{-(q_3-q_2)}+p_2 e^{-(q_1-q_3)}+p_3e^{-(q_2-q_1)}\ .
 \end{equation}

%---------------
\section{Weyl-Heisenberg covariant integral quantization}
%----------------
In this section, we give an overview of the quantization {that we use} in this paper to provide a quantum version of the classical Hamiltonian \eqref{condec2}. {It should be noted that the so-called canonical quantization (we recall it below) works sufficiently well for the basic observables and many typical Hamiltonians. {However,} it is neither the only possible quantization prescription nor it has to work well for more complex observables, in particular, when {the latter} admit some sort of singularities. The three canyons of the anisotropy potential, which narrow to the zero width, are in a sense singular and one may expect that the quantization should smooth out those singular features. For this reason, we employ a more general quantization framework which, apart from being well-suited for dealing with singular observables, possesses an attractive probabilistic interpretation. The fact that this quantization emphasizes the resemblance of the anisotropy Hamiltonian to the Toda system, which we prove in the next section, is another indication of the importance of the classical approximation that we have earlier proposed.}

For the sake of simplicity, we restrict the presentation   {of the method to the} Euclidean plane viewed as the phase space of the motion of a particle on the line,
\begin{equation}
\label{phaspace1}
\R^2= \{ (q,p)\, , \, q,p \in \R\}  \, .
\end{equation}
 We follow \cite{hbjpg2014,becugaro17_1,gazeau_intsympro18} where all details are given.
 Everyone is familiar with the basic procedure \cite{dirac82} which consists in the  replacements
 \begin{equation}
 \label{canquant}
 \begin{split}
\R^2\ni (q,p)  &\mapsto \ \mbox{self-adjoint}\ (\hq,\hp)\,, \quad [\hq,\hp] = \ii \hbar I\, ,\\
f(q,p) &\mapsto f(\hq,\hp) \mapsto (\mathrm{Sym}f)(\hq,\hp)\, ,
\end{split}
\end{equation}
where $\mathrm{Sym}$ stands for a ``certain" symmetrization, i.e. ordering. We  remind that $[\hq,\hp] = \ii \hbar I$ holds true with self-adjoint $\hq$, $\hp$, only if both have continuous spectrum $(-\infty,+\infty)$.
Now, we have $f(q) \mapsto f(\hq)$ (or $f(p) \mapsto f(\hp)$), which entails that the procedure does not cure singularities pre-existing in the classical model.  Also, the method is notoriously hard to implement as soon as the classical $f$ is not a polynomial, for instance the (singular!) angle function $f(q,p)= \arctan(p/q)$. Hence, we propose an alternative method which can get rid of these difficulties, specially those concerning ordering and singularities. The idea is to combine symmetry with resources of measure/integral calculus, where we can ignore points or lines to some extent. It can be viewed as an extension of the so-called Wigner-Weyl integral quantisation by combining the latter with a regularizing {weight function}, a kind of coarse-graining,  on the classical phase space, say
\begin{equation}
\label{distps}
(q,p) \mapsto \Pi(q,p)\,, \quad \mbox{with} \quad \Pi(0,0) = 1\, .
\end{equation}
While this function is not required to be a probability distribution,
this property is required for its symplectic Fourier transform, defined as
\begin{equation}
\label{symFourqp1}
 \mathfrak{F_s}[\Pi](q,p)= \int_{\mathbb{R}^2}e^{-\mathrm{i} (qp^{\prime}-q^{\prime}p)}\, \Pi(q^{\prime},p^{\prime})\,\frac{\mathrm{d}q^{\prime}\,\mathrm{d}p^{\prime}}{2\pi} \, ,
\end{equation}
for reasons which are explained in \cite{becugaro17_1,gazeau_intsympro18}. $\mathfrak{F_s}$  is involutive, $\mathfrak{F_s}\left[\mathfrak{F_s}[f]\right]=  f$ like its \textit{dual} defined as $\overline{\mathfrak{F_s}}[f](q,p)= \mathfrak{F_s}[f](-q,-p)$.

Let us now consider the operator
defined by the integral
\begin{equation}
\label{IWHT1}
\mathfrak{Q}_0 = \int_{\mathbb{R}^2} U(q,p) \, \Pi(q,p)\,
\frac{\mathrm{d}q\,\mathrm{d}p}{2\pi} \, ,
\end{equation}
where $U(q,p)$ is the Weyl unitary displacement operator, built from the self-adjoint  {$\hat q$ and $\hat p$} mentioned above,
\begin{equation}
\label{displacement}
U(q,p)= e^{\ii (p\hat q-q \hat p)}\, .
\end{equation}

The map $(q,p) \mapsto U(q,p)$ is a projective unitary irreducible representation of the abelian group $\R^2$. Then, by using Schur's Lemma,  the following resolution of the identity holds true:
\begin{equation}
\label{resunit2}
\int_{\R^2}  \mathfrak{Q}(q,p)\, \frac{\mathrm{d}q\,\mathrm{d}p}{2\pi} = I\,, \quad \mathfrak{Q}(q,p)= U(q,p)\mathfrak{Q}_0 U^{\dag}(q,p)\, .
\end{equation}

Equipped with one choice of $\Pi(q,p)$ or equivalently $\mathfrak{Q}_0$, one proceeds with the corresponding Weyl-Heisenberg covariant integral quantization,
\begin{equation}
\label{fAf}
 f(q,p) \mapsto A_f \equiv A^{\mathfrak{Q}_0}_f= \int_{\R^2} f(q,p)\, \mathfrak{Q}(q,p)\, \frac{\mathrm{d}q\,\mathrm{d}p}{2\pi} \, .
\end{equation}
The above quantization based on $\mathfrak{Q}_0$ is  possible if $\mathfrak{Q}_0$ is unit trace. Indeed, we have the relation
\begin{equation}
\label{WHtr}
\Pi(q,p) = \mathrm{Tr}\left(U(-q,-p)\mathfrak{Q}_0 \right)\, ,
\end{equation}
which can be interpreted as the \textit{Weyl-Heisenberg transform} of operator $\mathfrak{Q}_0$, Eq.\eqref{IWHT1} being viewed as the \textit{inverse Weyl-Heisenberg transform} of the distribution $\Pi(q,p)$. The unit trace condition is a consequence of  \eqref{WHtr} since  $1= \Pi(0,0) = \mathrm{Tr}\left(\mathfrak{Q}_0 \right)$.

In this context, the operator $\mathfrak{Q}_0$ is the quantum version (up to a constant) of the origin of the phase space, identified with the $2\pi\times$ Dirac distribution at the origin.
\begin{equation}
\label{quantdirac}
2\pi \delta(q,p) \mapsto A_{\delta}= \mathfrak{Q}_0\,.
\end{equation}
Translational covariance {means that the origin of the phase space in arbitrary. It} holds in the sense that the quantization of the translation of $f$ is unitarily equivalent to the quantization of $f$  as follows:
\begin{equation} \label{covtrans1}
U(q_0,p_0)\,A_f \,U^{\dag}(q_0,p_0)= A_{\mathcal{T}(q_0,p_0)f}\, , \quad \left(\mathcal{T}(q_0,p_0)f\right)(q,p):= f\left(q-q_0, p-p_0\right) \, .
\end{equation}
The properties of $U(q,p)$ allow us to establish an equivalent form of  the WH integral quantization:
 \begin{equation}
 \label{SIQ}
A_f= \int_{\mathbb{R}^2}  U(q,p)\,  \overline{\mathfrak{F_s}}[f](q,p)\, \Pi(q,p) \,\frac{\mathrm{d}q\,\mathrm{d}p}{2\pi}\,.
\end{equation}
There are several features  independent of the choice of the quantization ingredients $\mathfrak{Q}_0$ or $\Pi(q,p)$.
First, we have  the general important result:  if $\mathfrak{Q}_0$ is symmetric, i.e. $\overline{\Pi(-q,-p)}= \Pi(q,p)$,  a real function $f(q,p)$ is mapped to a symmetric operator $A_f$. Moreover, if  $\mathfrak{Q}_0$ is a non-negative operator, i.e., is a density operator, then  a real semi-bounded  function $f(q,p)$ is mapped to a self-adjoint operator $A_f$ through the Friedrich extension of its associated semi-bounded quadratic form.
Next, the canonical commutation rule is preserved
\begin{equation}
\label{AqAp}
q\mapsto A_q = \hat q + c_0\, , \quad p\mapsto A_p= \hat p+d_0\,, \quad c_0,d_0\in \mathbb{R}\, ,  \rightarrow \left[A_q,A_p\right]= \mathrm{i} I\, .
\end{equation}
For the kinetic energy we have the following formula
\begin{equation}
p^2\mapsto A_{p^2}= \hat p^2 + e_1\,\hat p + e_0\, , \quad e_0, e_1 \in \mathbb{R}\, .
\end{equation}
The constants $c_0$, $d_0$, $e_0$, $e_1$ appearing in the above can be made equal to $0$ with a suitable choice of $\mathfrak{Q}_0$ and thus, we may re-establish the well-known results of the canonical quantization. The quantization of the dilatation  operator yields:
\begin{equation}
qp\mapsto A_{qp} = A_q\,A_p + \mathrm{i} f_0\, , \quad  f_0\in \mathbb{R}\, .
\end{equation}
This operator can be brought to the self-adjoint dilation operator $(\hat q\hat p + \hat p\hat q)/2$ again with a suitable choice of $\mathfrak{Q}_0$.

A potential energy becomes the multiplication operator in the position representation
\begin{equation}
\label{Iqvq}
V(q)\mapsto A_{V(q)} = \mathfrak{V}(\hat q)\, , \quad \mathfrak{V}(\hat q)= \frac{1}{\sqrt{2\pi}}\,V\ast \overline{\mathcal{F}}[\Pi(0,\cdot)](\hat q)\,
\end{equation}
where $ \overline{\mathcal{F}}$ is the inverse $1$-$d$ Fourier transform, and $f\ast g(x)= \int_{\mathbb{R}}\mathrm{d} t\, f(x-t)\,g(t)$. Such a convolution formula can be of crucial importance when it is needed to smooth classical singularities or modify in a suitable way the strengths of some potentials as will be shown in the sequel.

Finally, if $F(q,p)\equiv h(p)$ is a function of $p$ only, then $A_h$ depends on $\hat p$ only
\begin{equation}
\label{Iqvp}
A_h= \frac{1}{\sqrt{2\pi}}\,h\ast \overline{\mathcal{F}}[\Pi(\cdot,0)](\hat p)\, .
\end{equation}
In the next section, we choose separable Gaussian distributions:
\begin{equation}
\Pi(q,p) = e^{-\frac{q^2}{2\sigma^2}}\, e^{-\frac{p^2}{2\tau^2}}.
\end{equation}
They
 yield simple formulae with familiar probabilistic content.  Standard coherent state (or Berezin or anti-Wick)  quantization corresponds to the particular values $\sigma=\sqrt{2}=\tau$.
The limit Weyl-Wigner case holds as  the widths $\sigma$ and $\tau$ are infinite (Weyl-Wigner is singular in this respect).

%---------------------------------
\section{Quantization of the potential}
%----------------------------------

\subsection{General result}
We apply the integral quantization method described in the previous section to the anisotropy potential. 
{For the quantization of the two-dimensional anisotropy potential \eqref{b9pot} we employ the separable Gaussian distributions mentioned in the previous section:
\begin{equation}
\Pi(\beta_+,p_+;\beta_-,p_-)=e^{-\frac{\beta_+^2}{2\sigma^2_+}}\,e^{-\frac{\beta_-^2}{2\sigma^2_-}}\,e^{-\frac{p_+^2}{2\tau_+^2}}\,e^{-\frac{p_-^2}{2\tau_-^2}},
\end{equation}
where we use the formula:
\begin{equation}
e^{\alpha q}\mapsto {A}_{e^{\alpha q}}=\frac{1}{\sigma\sqrt{2\pi}}\int_{-\infty}^{\infty}e^{-\frac{(q-y)^2}{2\sigma^2}} e^{\alpha y} \,\textrm{d}y=e^{\frac{\alpha^2}{2\sigma^2}}e^{\alpha q}.
\end{equation}
We obtain}
\begin{align}
V(\beta_+,\beta_{-})\mapsto {A}_{V(\beta_+,\beta_{-})}=\frac13&\left(2D_+^4 D_-^{12} e^{4\beta_+}\cosh 4\sqrt{3}\beta_--4
D_+ D_-^3 e^{-2\beta_+}\cosh 2\sqrt{3}\beta_-\right.
\nonumber\\ &\left.+D_+^{16} e^{-8\beta_+}-2D_+^4 e^{4\beta_+}\right) +1,\label{qpot9}
\end{align}
where $D_+= e^{\frac{2}{\sigma^2_+} }$ and $D_-=  e^{\frac{2}{\sigma^2_-} }$. The classical anisotropy potential $V(\beta_+,\beta_{-})$ is recovered for $D_+=D_-=1$ (or, $\sigma_+,\ \sigma_-\rightarrow \infty$).

Figure \ref{figure2} (left plot) shows the quantum potential \eqref{qpot9} for sample values of $D_+$ and $D_-$. The classical escape canyons are absent and the potential is fully confining. Moreover, it is anisotropic in the variables $\beta_+$ and $\beta_-$, and the position of its minimum is, in comparison to the classical case, shifted from $(0,0)$ to $( \beta_0,0)$, where the value $\beta_0$ satisfies the following equation:
\begin{equation}\label{beta0}
  -D_+^{16} e^{-8 \beta_0}+D_+ D_-^3 e^{-2 \beta_0}-D_+^4 e^{4 \beta_0}+D_+^4 D_-^{12} e^{4 \beta_0}=0,
\end{equation}
derived from condition $\partial{A}_{V(\beta_+,\beta_{-})}/\partial \beta_+=0$ (condition $\partial{A}_{V(\beta_+,\beta_{-})}/\partial \beta_-=0$ is trivially fulfilled).
The minimal value of the potential is shifted from  $0$ to the value:
\begin{equation}\label{qpot00}
 {A}_{V(\beta_+,\beta_{-})}(\beta_0,0)=\frac{1}{3} e^{-8 \beta_0} \left(D_+^{16}-4 D_+ D_-^3 e^{6 \beta_0}+3 e^{8 \beta_0}+2 D_+^4 \left(-1+D_-^{12}\right) e^{12 \beta_0}\right).
\end{equation}

\subsection{Minimum and isotropy preservation}
We want our quantization procedure to preserve the basic properties  of the classical potential by requiring (i) the isotropy around its minimum and (ii) assumes the position of the minimum at  the point $(0,0)$.

Expansion of the potential \eqref{qpot9} around the point $(0,0)$\footnote{Expanding around $(\beta_0,0)$ yields similar results but the presentation is far more sophisticated.} reads:
\begin{align}\label{vexpanded}
{A}_{V(\beta_+,\beta_{-})}&\approx \frac13 \left[ D_+^{16}
-2 D_+^4+2D_+^4 D_-^{12}  -4D_+ D_-^3
-8\left( D_+^4 +D_+^{16}- D_+ D_-^3 - D_+^4 D_-^{12} \right) \beta_+\right. \nonumber\\
&\left.+8\left(4 D_+^{16} -2 D_+^4 -  D_+ D_-^{3}+2 D_+^4 D_-^{12} \right) \beta_+^2+24\left(2 D_+^4 D_-^{12} - D_+ D_-^3 \right) \beta_-^2\right] +1.
\end{align}
The coefficients in front of $\beta_+^2$ and $\beta_-^2$ are equal iff
$$
4 D_+^{16} -2 D_+^4 -  D_+ D_-^{3}+2 D_+^4 D_-^{12}=3\left(2 D_+^4 D_-^{12} - D_+ D_-^3 \right).
$$
This condition may be rewritten as:
\begin{equation}\label{isocon0}
  \left(D_+^{4}-D_+D_-^{3}\right)\left[2\left(D_+^{4}+D_+D_-^{3}\right)\left(D_+^{8}+(D_+D_-^{3})^2\right)-1\right]=0.
\end{equation}
For values of   $D_+, D_-\ge 1$ (condition fulfilled by exponents) it has  only one unique solution:
\begin{equation}\label{isocon}
 D_+=D_-:=D.
\end{equation}
The condition (\ref{isocon}) also implies the vanishing of the coefficient in front of $\beta_+$ in Eq. (\ref{vexpanded}). \\
{This is consistent with the Eq. \eqref{beta0}, which for $\beta_0=0$ can be rearranged as:}
\begin{equation}
  \left(D_+^{4}-D_+D_-^{3}\right)\left[\left(D_+^{4}+D_+D_-^{3}\right)\left(D_+^{8}+(D_+D_-^{3})^2\right)+1\right] =0,
\end{equation}
{which  again for positive values of $D_+$ and $D_-$ has only one unique  solution $D_+=D_-$, the same as the one for symmetry preservation}. Thus, there is no shift in the position of the minimum.

\begin{figure}[!ht]
\begin{tabular}{cc}
\includegraphics[width=0.35\textwidth]{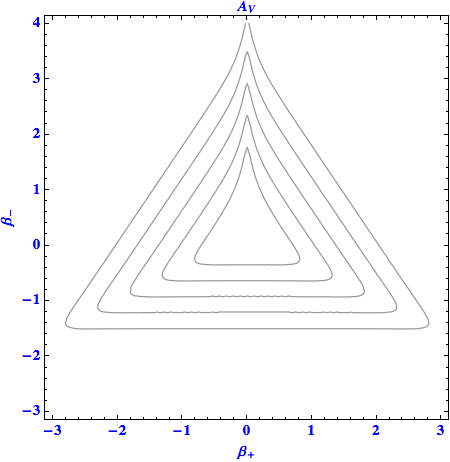}
\includegraphics[width=0.35\textwidth]{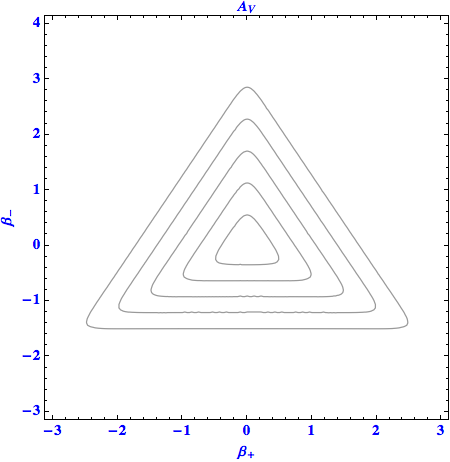}
\end{tabular}
\caption{On the left contour plot of the quantized Bianchi IX potential near its minimum, for sample values $D_+=1.2$, $D_-=1.005$. On the right contour plot  of the symmetric quantized  potential near its minimum, $D_+=D_-=1.2$. }
\label{figure2}
\end{figure}

The full quantized Bianchi IX potential reads now:
\begin{align}\nonumber
 {A}_{V(\beta_+,\beta_{-})}=\frac13&\left(2D^{16}e^{4\beta_+}\cosh 4\sqrt{3}\beta_- -4 D^{4}e^{-2\beta_+}\cosh 2\sqrt{3}\beta_-\right.\\\label{isopot}&+D^{16}e^{-8\beta_+}
 \left.-2D^{4}e^{4\beta_+}\right)+1.
\end{align}
It is shown in Fig. \ref{figure2} (right plot). One may verify that it is invariant with respect to the rotations by $2\pi/3$ and $4\pi/3$, and thus, the $\mathsf{C}_{3v}$ symmetry is preserved in the full plane.

\subsection{ Underlying Toda system}
The quantized potential \eqref{isopot}, in full analogy to the classical potential in Eq. (\ref{split}), may be written as a difference between two Toda potentials ({recall} that $V_p$ is negative):
\begin{align}\label{isopot1}
 {A}_{V(\beta_+,\beta_{-})}=&D^{16}V_T+D^{4}V_p+1,
\end{align}
where the coefficients $D^{16}$ and $D^{4}$ strengthen the dynamical role of $V_T$ as $D>1$. %------------------------------------------
%------------------------------------------
We are interested in verifying whether the following
\begin{equation}
\left|\frac{{V}_{p}}{{V}_{T}}\right|=\left|\frac{2XY+1}{2Y^2+X^2/2-1} \right|,
\end{equation}
is bounded from above, where
\begin{align}
X= &e^{-6\beta+}\ge 0,\\
Y=&\cosh2\sqrt{3}\beta_-\ge1.
\end{align}
Direct computation shows that in the open domain $(X,Y)\in\,]\,0,\infty[ \,\times \,]\,1,\infty\,[$ the ratio $\left|\frac{{V}_{p}}{{V}_{T}}\right|(X,Y)$ does not have any critical points. On the boundary $Y=1$ it has one maximum at $X=1$,  with $\left|\frac{{V}_{p}}{{V}_{T}}\right|{(1,1)}=2$, which is therefore a unique global maximum. This proves that for exponential weights $D>1$, the exact quantum potential may be viewed as a  perturbed Toda potential $D^{16}V_T$.

%-----------------------------------
\section{Discussion}
%--------------------------------------
{The Bianchi IX model plays  an essential role in our understanding of the near-singularity dynamics of the universe. The classical Hamiltonian for this model involves an elaborate  two-dimensional anisotropy potential. Therefore, the quantizations of the model found in the literature often rely on one of the two approximations to the anisotropy potential: the steep-wall or the harmonic approximation. They enable to analytically determine  the respective eigenstates and eigenvalues and apply adiabatic or nonadiabatic approaches to study the quantum dynamics of that model.}

{In this work, we developed a new proposal for approximating the anisotropic part of the quantum Hamiltonian with the integrable 3-particle Toda system. It  approximates the three exponential walls while smoothing out the problematic canyons. This result  brings new possibilities to studies  of the quantum dynamics of mixmaster. The periodic 3-body Toda system has been analysed both on the classical and quantum levels. In the literature, one may find ways to construct classical solutions \cite{Flaschka,KacMoerbeke} as well as the corresponding quantum eigenfunctions and eigenvalues \cite{Gutzwiller,VanMoerbeke}.  Thus, we provide an analytically solvable approximation to the anisotropic potential of the Bianchi IX model not only in the IR and UV limits, given by the harmonic and steep-wall approximation respectively, but also in the vast and unexplored  in-between region.}

%%%%%%%%%%%%%%%%%%%%%%%%%%%%%%%%%%%%%%%%%%%%%%%%%%%%%%%%%%%%%%%%%%%%%%%%%
\begin{acknowledgments}
The authors are grateful  to
Alfredo Miguel Ozorio de Almeida (Centro Brasileiro de Pesquisas F\'{\i}sicas, Rio de Janeiro (RJ)) for  pointing out the paper \cite{berry76} and the possible relevance  of the Toda system for the Bianchi IX model.
\end{acknowledgments}

\end{document}